# Giant reversible nanoscale piezoresistance at room temperature in $Sr_2IrO_4$ thin films

*Neus Domingo,[1,*] Laura López-Mir,[1,2] Marcos Paradinas,[2] Vaclav Holy,[3] Jakuv Zelezny,[4] Di Yi,[5] Siriyara J. Suresha,[6] Jian Liu,[7] Claudy Rayan-Serrao,[5] Ramamoorthy Ramesh,[5,6,7] Carmen Ocal,[2] Xavi Martí,[1,4,8♠] Gustau Catalan.[1,9,♣]*

[1] ICN2-Institut Català de Nanociència i Nanotecnologia, Campus Universitat Autònoma de Barcelona, 08193 Bellaterra, Spain

[2] ICMAB, Institut de Ciència de Materials de Barcelona (CSIC), Campus Universitat Autònoma de Barcelona, 08193 Bellaterra, Spain

[3] Department of Condensed Matter Physics, Faculty of Mathematics and Physics, Charles University, 12116 Praha 2, Czech Republic

[4] Institute of Physics ASCR, v.v.i., Cukrovarnická 10, 162 53 Praha 6, Czech Republic

[5] Department of Materials Science and Engineering, University of California, Berkeley, California 94720, USA

[6] Materials Sciences Division, Lawrence Berkeley National Laboratory, Berkeley, California 94720, USA

[7] Department of Physics, University of California, Berkeley, California 94720, USA

[8] IGS Research, C/ La Coma, Nave 8, 43140 La Pobla de Mafumet (Tarragona), Spain

[9] ICREA - Institució Catalana de Recerca i Estudis Avançats, Barcelona

KEYWORDS: giant piezoresistance, Mott insulators, piezotronics, AFM, uniaxial pressure, iridates.

ABSTRACT:

Layered iridates have been the subject of intense scrutiny on account of their unusually strong spin-orbit coupling, which opens up a narrow gap in a material that would otherwise be a metal. This insulating state is very sensitive to external perturbations. Here, we show that vertical compression at the nanoscale, delivered using the tip of a standard scanning probe microscope, is capable of inducing a five orders of magnitude change in the room temperature resistivity of $Sr_2IrO_4$. The extreme sensitivity of the electronic structure to anisotropic deformations opens up a new angle of interest on this material, and the giant and fully reversible perpendicular piezoresistance makes iridates a promising material for room temperature piezotronic devices.

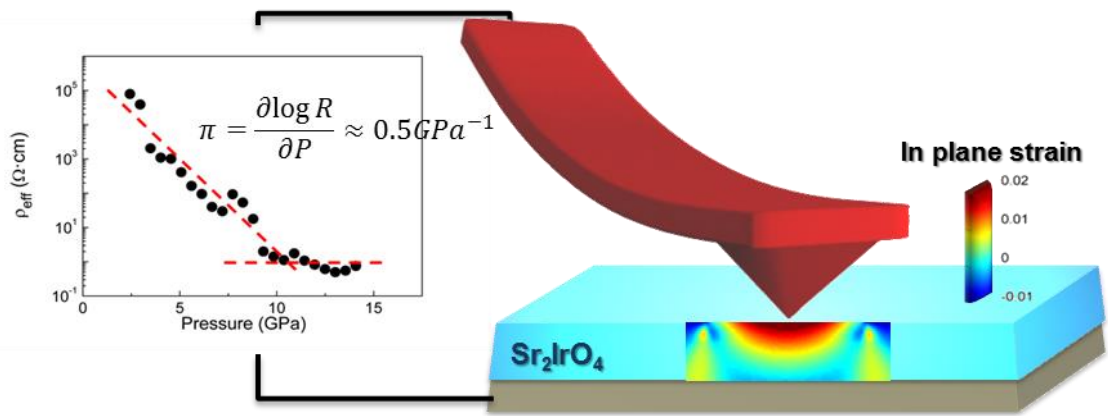

Piezoresistance defines the change in electrical resistance of a material as a function of a mechanically induced deformation. Originally, its main use was in deformation sensors (strain gauges), but it is now moving into the center stage of the electronic industry as the basis for new transistor concepts such as Piezoelectronic Transistors (or PET),[1,2] aiming to circumvent the gate-voltage bottleneck in transistor miniaturization.[3]

The term "giant piezoresistance" was used by He and Yang[4] to describe the 40 times bigger piezoresistance of Si nanowires compared to bulk silicon.[5-7] While the nature and origin of the effect in Si nanowires still remains controversial,[8] it is generally accepted that the increase in conductivity of semiconductors such as silicon or germanium under strain is due to an enhanced mobility of their carriers.[9,10] An alternative route towards giant piezoresistance does not change the mobility of the carriers but their number density in materials where the band gap is strain dependent. Archetypal examples of giant piezoresistance through this mechanism are rare-earth selenides,[1] for which the proximity to a metal-insulator transition renders resistivity extremely sensitive to mechanical deformations.

Oxides are a fertile ground in the search for novel or enhanced electronic phenomena, with notable examples including high temperature superconductivity, colossal magnetoresistance, and room temperature magnetoelectric multiferroicity.[11] Some perovskite oxides, such as manganites and vanadates,[12] are also known to exhibit very large piezoresistance,[13,14] but unfortunately their features are usually optimal only in the immediate vicinity of temperature-driven metal-insulator transitions. In this context, strontium iridate $Sr_2IrO_4$ (SIO) appears as a promising material where the strong competition of spin-orbit coupling, Coulomb on-site repulsion, and crystal field opens a conduction gap in a material that would otherwise be a metal.[12,15-18] Chemical substitutions,[19] tuning of the Ruddlesden-Popper series' dimensionality,[20] interface exchange-coupling,[21] magnetic fields,[19] etc., can all unbalance the subtle equilibrium of competing interactions and therefore alter the electronic transport. SIO has been realized as epitaxial thin film recently,[21-23] and substrate-induced strain has been observed to cause a substantial modification of its transport activation energy[22] and optical bandgap,[24] which suggests that SIO might be a good piezoresistive material. In this paper, we report five orders of magnitude and fully reversible piezoresistive change at room temperature in thin films of SIO induced by pressure perpendicular to the surface.

Piezoresistive characterization is typically performed by applying uniform strain to the samples fixing their extremes in microelectromechanical systems.[25] Here we propose a different method, based on the use of Atomic Force Microscopy, which not only allows us to deliver high pressures with tiny forces, but moreover enables high frequency sampling and measurement of local piezoresistance with nanoscopic lateral resolutions. The pressure is applied by controlling the load exerted by a conductive tip of an atomic force microscope (AFM) with nanometric radius, whilst simultaneously measuring the current across the film by connecting the tip to an amplifier. The tip thus acts as both anvil cell and top electrode, and the small contact area enables delivering very high pressures in the GPa range using only modest forces of the order of $\mu$N, making this a very convenient bench-top tool for high pressure experiments. The measured piezoresistance (relative increase in resistance divided by applied pressure) is $10000 \cdot 10^{-11}$ Pa$^{-1}$; this is larger than the giant piezoresistance of silicon nanowires[4] and comparable to that of rare earth selenides,[1] marking the emergence of Sr$_2$IrO$_4$ as a viable oxide for piezotronics.

RESULTS AND DISCUSSION

The sample under study is a SIO 6 nm-thick film, grown on 12 nm-thick La$_{0.7}$Sr$_{0.3}$MnO$_3$ (LSMO) metallic bottom electrode, on top of a SrTiO$_3$ (001) (STO) substrate[21] (see experimental section and Electronic Supplementary Information for sample preparation and characterization details). Local transport measurements have been performed by placing a AFM tip in direct contact with the sample under controlled load, i.e. by using a normal force feedback, and collecting the current flowing between top electrode (conducting tip) and bottom electrode (metallic LSMO layer) as a function of voltage, $I(V_{bias})$, for different mechanical loads kept below plastic deformation of the material (Figure 1a). Acting both as top electrode and pressure anvil, the AFM tip collects the current perpendicular to the plane (CPP) flowing across the SIO layer. The sharpness of the AFM tips ensures that high uniaxial stresses can be delivered with only modest forces. The measurement is sketched in Figure1a and the qualities of the epitaxial SIO film as well as that of the SIO/LSMO interface are seen in the scanning transmission electron microscopy (STEM) image in Figure 1b. Continuum elastic calculations for the tip-induced deformation using the Hertzian model (see the in-plane and out of plane strain distributions shown in Figure 1c and 1d, respectively) renders that 10 µN of force translates into 10.1 GPa of stress. Combined with and DFT-calculated elastic constants (see Electronic Supplementary materials) the tip-induced vertical compression is calculated to reach up to -7%. This severe reduction of the tetragonality dramatically modifies the conductivity.

Local transport measurements (Figure 2a) show a rapid increase of the $I(V_{bias})$ slope (i.e., a conductance rise) as the mechanical load exerted by a conductive AFM tip is increased. Time-dependent current data, collected at a constant force and bias voltage, showed no evidence of transient behavior, and the surface topography was intact after measurements, indicating that increase in current is not due to piercing of the SIO film. Figure 2b shows the total effective resistance (R) as a function of applied load (F) obtained from the slope of the $I(V_{bias})$ in the linear regime, i.e., at low voltages, for different sets of experiments. All the data obtained fall into a single curve displaying an exponential decay of the resistance (note the logarithmic R scale) as a function of the applied force and independently of the surface point or the environmental conditions. Beyond F*~10 µN and up to our maximum recorded force of 25 µN, the resistance reaches a saturation value becoming force independent.

The 3D representation of the electronic current as a function of both, voltage and applied force, $I(V_{bias}, F)$, is depicted in Figure 2c (see 3D modes description in Experimental Section). The increase in conductance for increasing load is evident from the gap reduction seen as a narrowing of the nearly flat region (intermediate color in the I scale) at low $V_{bias}$. The use of mechanically robust and stiff diamond-coated tips with $k = 40$ N/m allows us modeling the elastic deformation of the film under the tip, and thus converting resistance into resistivity. The tip radius determination and its mechanical stability were verified using high resolution SEM images that show no apex deterioration and a radius $r_{tip} = 150$ nm both before and after the experiment (inset in Figure 2d). The force versus displacement curves of the AFM cantilever were reversible, indicating the absence of plastic deformation, invariant tip-sample effective spring constant and negligible tip-sample adhesion (see Electronic Supplementary Information), thus validating the applicability of the Hertzian elastic model for calculating tip-induced deformation.[26, 27] The Hertz effective contact radius $a$ as a function of applied force is:

$$a = \left(\frac{3}{4}\frac{F \cdot r_{tip}}{E^*}\right)^{1/3} \qquad (1)$$

Where $E^*$ is the Young elastic modulus and $r_{tip}$ is taken from the SEM images. With these values, we can determine the mean contact pressure exerted on a column of SIO, as:

$$P_m = \left(\frac{4E^*}{3\pi}\right)\frac{a}{r_{tip}} \qquad (2)$$

and estimate an effective resistivity ($\rho_{eff}$) for different loading conditions on the basis of the normalization of the total resistance ($R$) by the effective tip-sample contact area:

$$\rho_{eff}(P_m) = \left(R \cdot \frac{\pi r_{tip}^2}{d - a^2/r_{tip}}\right)^{1/3} \qquad (3)$$

where we also consider the reduction on the film thickness *d* due to the deformation depth $\delta = \frac{a^2}{r_{tip}}$. The obtained effective resistivity as a function of the mean contact pressure is plotted in Figure 2d and show an outstanding five orders-of-magnitude decrease in resistivity. Notice that these calculations already take into account the tip-sample contact area and thus the increase in conductivity is not due to any increase of contact area or any other geometric effect. We also notice that the high-pressure resistivity is much smaller than the 4-probe resistivity of bare SIO thin films ($10^3$ Ω·cm at room temperature).[28] This indicates that the piezoresistive effects cannot be attributed to changes in the tip, because otherwise the resistivity would saturate at the resistivity of the highest-resistance element in the circuit, which in the absence of piezoresistance would be $10^3$ Ω·cm. Moreover, it shows that the perpendicular deformation induced by the tip is more effective at modulating resistivity than the in-plane deformation induced by epitaxy.

The robustness and reversibility of the piezoresistive effect have been tested by introducing a measurement protocol whereby the AFM repeatedly scans the same 125 nm-long line on the SIO film surface over 500 times, shifting back and forth the force between alternate set points (0.5 μN and 2.0 μN) on each consecutive scan line, while measuring simultaneously measuring the conductivity with a constant bias voltage of 200 mV (Figure 3). This protocol further rules out the possibility of resistive switch being dominated by undetected adsorbate overlayers: any adsorbates, if initially present, would be removed after the first scan. Meanwhile, the simultaneously monitored surface topography, shown in Figure 3a, did not change during the entire test. Figure 3d displays the average current per line-scan for each alternating load. The data reveal two persistent CPP states, separated by more than two-orders of magnitude (on-off resistance ratio of 250 for a force ratio of only 4), reversibly and accurately reproduced still after 500 scans.

The average contact stress corresponding to each of the two set points is 3.4 GPa and 6 GPa, respectively; so the conductive piezoresistance coefficient ($\pi_l^\sigma = \frac{1}{X}\frac{\Delta\sigma}{\sigma_0}$, where X is the stress and $\sigma_0$ is the conductivity under the minimum stress) is $10^{-7}$ Pa$^{-1}$, about 3 times larger than that of Si nanowires with giant piezoresistance.[4] The logarithmic coefficient is $\pi_{log} = \frac{\partial \log R}{\partial P} \approx 0.5\ GPa^{-1}$, to be compared to $\pi_{log} \approx 1\ GPa^{-1}$ for SmSe thin films of similar thickness.[1] Meanwhile, the piezoresistive gauge factor, which is the change in resistance normalized by *strain* instead of stress, $G = \frac{1}{\Delta\epsilon}\frac{\Delta R}{R_0}$ can be estimated using the out-of-plane deformations averaged over the cylinder directly under the contact area ($\epsilon_{zz}$ = -0.0213 and $\epsilon_{zz}$ = -0.0126 for 2 µN and 0.5 µN respectively), yielding a value G = 25000, which is bigger than the giant piezoresistive ratio of oxide cobalates[29] (G = 7000) and bigger even than that of graphene (G=18000).[30, 31] The piezoresistive and gauge coefficients or SIO are therefore comparable to or exceeding those of the best piezoresistive materials. In addition, as oxides that can be grown epitaxially on perovskites, iridates are naturally compatible with the piezoelectric elements in piezoelectronic transistors.[32]

We now examine the origin of the enormous change in electrical resistivity. The elastic calculations indicate that ~10 µN forces translate into pressures of ~10 GPa. Under hydrostatic conditions, such pressures are sufficient to cause a substantial closure of the conduction band gap.[15] Hydrostatic pressure experiments reduce all inter-atomic distances, which should naturally increase bandwidth and decrease bandgap. However, due to the higher compressibility of the in-plane bond, hydrostatic pressure also results in an increased tetragonality of SIO.[15] In such experiments, the bandgap decreases from 60 to 30 meV at a pressure of c.a. 15 GPa, but saturates thereafter and metallic state is never reached even up to 100 GPa. Meanwhile, epitaxial tensile strain expands the in-plane lattice parameter and, via Poisson's ratio, contracts the out-of-plane lattice parameter, resulting in a decrease of the tetragonality of thin films. Experimentally, a 1% decrease in tetragonality is correlated with a reduction in the transport gap from 200 meV to 50 meV.[22] This result already indicates that anisotropic deformations are more efficient than hydrostatic pressure in modulating the bandgap. As a matter of fact, electric transport in iridates is known to be highly anisotropic: single crystal experiments show that resistivity along the c-axis is at least two orders of magnitude larger than along the a and b axis[19], consistent with the larger Ir-Ir distance brought about by the rocksalt-like SrO intercalated layers sandwitched between perovskite-like layers in the Ruddlesen-Popper structure of SIO. It is therefore natural to expect that a reduction of out-of-plane interlayer distances will reduce the large out-of-plane resistivity. However, ascribing the enormous change in electrical resistivity solely to a reduction of the interplanar distances can't be straightforwardly reconciled with the fact that, if the in-plane resistivity marks the lower limit, then one would expect one two orders of magnitude change resistance at most, instead of the observed 5 orders. It is therefore plausible that, together with the lattice contraction, there is also a straightening of in-plane Ir-O-Ir bonding angles which further contributes to the reduction of bandgap as calculated in Ref. 33.

Such a constriction of the bandgap could drive the large observed changes in electrical transport but the confirmation of a metallic ground-state occurring after the large strains localized below the probe tip remains a topic for subsequent studies. In the present nanoscale experiments, the giant decrease of resistance is obtained under a deformation field where the dominant term is vertical compression, with a force of 10 μN causing a reduction of tetragonality by 7% (Figure 1c and 1d) correlated with the 100000% decrease in resistivity, suggesting that the perpendicular resistivity is predominantly controlled by inter-atomic distances along the c-axis.

CONCLUSIONS:

Epitaxial films of the Mott insulator SIO show a pressure-induced, fully reversible, orders of magnitude nanoscale modulation of CPP transport –giant piezoresistance– at room temperature. This can be induced with modest forces in the µN range thanks to the use of nanometric (~150 nm radius) conducting scanning probes to deliver the stress while simultaneously collecting the current.

In SIO, the perpendicular piezoresistive effect is facilitated by strong correlation effects within a highly anisotropic lattice. The layered structure of SIO renders the system very sensitive to perpendicular deformations, leading to piezoresistive and gauge coefficients comparable to or exceeding those of the best piezoresistive materials; the 5 order of magnitude change in resistance under 10GPa of vertical pressure is in fact the largest reported room temperature perpendicular piezoresistance for any oxide, and this opens up a new line of research and potential applications for layered iridates.

As oxides that can be epitaxially grown on perovskites, iridates are also naturally compatible with the piezoelectric ceramics used as actuator elements in piezoelectronic transistors,[32] making them very attractive for such devices. In addition, spin may also be exploited separately to manipulate the transport properties below the Néel temperature,[19, 33] while strain manages stronger variations of the bandwidth and thus larger resistance changes at room temperature.

EXPERIMENTAL SECTION:

**Samples preparation and characterization:** The heterostructure is deposited by pulsed laser deposition (PLD) assisted by reflection high-energy electron diffraction (RHEED) as detailed in previous work.[21] LSMO and SIO were grown epitaxial on the STO substrates from the stoichiometric targets at a laser energy density of ~1.5 J/cm$^2$ and a repetition rate of 1 Hz. LSMO is grown at 700 °C and 150 mTorr partial oxygen pressure while SIO is grown at 800 °C and 1 mTorr. After the layer by layer growth, the samples were cooled to room temperature in 760 Torr oxygen ambient at a rate of 5°C. Cross-section scanning transmission electron microscopy (STEM) image of the LSMO/SIO heterostructure (Figure 1b) indicates an atomically sharp interface between the LSMO and SIO, and the smooth surface of the heterostructure is revealed by atomic force microscopy (AFM) images as shown in Figure 1c, showing a root-mean square roughness below one unit cell (see Electronic Supplementary Information for more details on sample characterization).

**Transport measurements by AFM:** Transport measurements were performed locally by means of conductive AFM (C-AFM). The experimental setup consist of the sample and tip placed in series resulting in a capacitor where the tip is the top electrode and the LSMO thin film substrate between SIO and STO is the bottom electrode. Conductivity measurements through the film in the form of I(V$_{bias}$) curves, I(t) curves and C-AFM images at selected tip loads have been performed using an MFP3D Asylum AFM by means of ORCA module that consists of a specially-designed cantilever holder that encloses two different amplifiers with gains of 10$^6$ and 10$^9$ V/A which allow us to record currents from 1 pA to 10 μA. The conductive tip mounted in the cantilever holder is grounded and it is used as a probe to do the measurement, while the voltage is applied directly to the sample. Any current will flow from the bottom electrode of the sample, through the sample to the tip and finally to the current amplifier. This configuration allows us to apply compressive force directly onto the thin film while simultaneously recording the current-voltage characteristics. I(V$_{bias}$) curves are swept on specific points of a selected area of the sample and with a prefixed load controlled by the Voltage Set Point ($V_{sp}$). The bias voltage we applied ranges from 25 mV to 1 V, thus resistance values can be obtained from the linear inverse slope of the $I(V_{bias})$ curves around the origin. Applying a constant voltage for a period of time also gives information about the current stability. Different types of tips from Nanosensors were used in the experiments: NCH Pt ($k$ = 40 N/m, PtIr coating, nominal tip radius < 40 nm), FM PtSi ($k$ = 2 N/m, PtSi coating, nominal tip radius < 15 nm), CDT – NCH and CDT – FMR ($k$ = 40 N/m and $k$ = 6 N/m respectively, doped diamond coating, nominal tip radius between 100 nm and 200 nm). Metal-coated conductive AFM tips show higher levels of conductivity when compared with doped diamond coated AFM tips as observed in Figure 2 (explicit comparison is shown in Electronic Supplementary Information), but instead, the last ones are more mechanically robust and their geometry, as determined by SEM measurements (Figure 2), remains completely stable during the whole measurement at the highest applied loads.

**3D conductive maps:** The so-called 3D maps[34, 35] were performed using a commercial AFM microscope and software from Nanotech.[36] These modes rely on the imaging advantages of AFM to simultaneously measure one or more given magnitudes as a function of other two: $f_i(x_1, x_2)$, where the fast ($x_1$) and slow ($x_2$) scan directions do not necessarily correspond to longitudinal dimensions. The result is a 3D plot in which the color scale represents the magnitude $f_i$ and the horizontal ($x_1$) and vertical ($x_2$) axes represent the two magnitudes chosen as variables. In the present case, the applied normal force ($f_1$ = F) and the current ($f_2$ = I) between tip and sample, were monitored as a function of the tip–sample voltage ($x_1 = V_{bias}$) and vertical piezo-displacement ($x_2 = z$), obtaining $F(V_{bias}, z)$ and $I(V_{bias}, z)$ plots. Provided the normal force versus z can be determined from vertical profiles in $F(V_{bias}, z)$, a new $I(V_{bias}, F)$ plot can be easily constructed. The resistance is evaluated for each force from the slope of the corresponding $I(V_{bias})$ curve (horizontal profile) of this plot. In this case, the bias voltage was applied to the tip and the current through the sample was measured by an external current amplifier (SR570).

**Electronic Supplementary Information:**

Electronic Supplementary information provides results of: i) sample synthesis and characterization ii) stability of surface isolating state after pressure and influence of tip coating material. iii) calculation of effective resistivity; tip-sample mechanical coupling and geometrical effects, iv) calculation of elastic constant and v) simulation of uniaxial stress. This material is available free of charge via the Internet at http://pubs.acs.org

**Corresponding authors:**

\* E-mail: neus.domingo@cin2.es

♠ E-mail: xavi.marti@igsresearch.com

♣ E-mail: gustau.catalan@cin2.es

**Author Contributions:**

Sample fabrication: J.L., D.Y. R.R.; Scanning Transmission Electron Microscopy: S.S; Experiments and data analysis: N.D., L.L., C.O., M.P.; Theory and elastic simulations: J.Z., T.J., V.H.; Manuscript writing and project planning: N.D., G.C., T.J., L.L, X.M.

**Funding Sources**

N.D wants to acknowledge the Spanish Ministerio de Ciencia e Innovación for a Ramon y Cajal research grant RYC-2010-06365. X.M. acknowledges the Grant Agency of the Czech Republic No. P204/11/P339. G.C. acknowledges ERC Starting Grant 308023. T.J. acknowledges ERC Advanced Grant 268066 and Praemium Academiae of the Academy of Sciences of the Czech Republic. Di Yi was sponsored by the National Science Foundation through the Penn State MRSEC. J.L. is supported by the Director, Office of Science, Office of Basic Energy Sciences, Materials Sciences and Engineering Division, of the U.S. Department of Energy under Contract No. DE-AC02-05CH11231 through the Quantum Material program in the Materials Sciences Division of Lawrence Berkeley National Laboratory. Financial support has been obtained under projects from the Spanish Ministerio de Economía y Competitividad under projects MAT2010-17771, MAT2010-20020, MAT2013-47869-C4-1-P, FIS2013-48668-C2-1-P and NANOSELECT CSD2007-00041 and Severo Ochoa Excellence Programme 2013-0295, and the Generalitat de Catalunya under projects 2014 SGR 733 and 2014 SGR 1216.

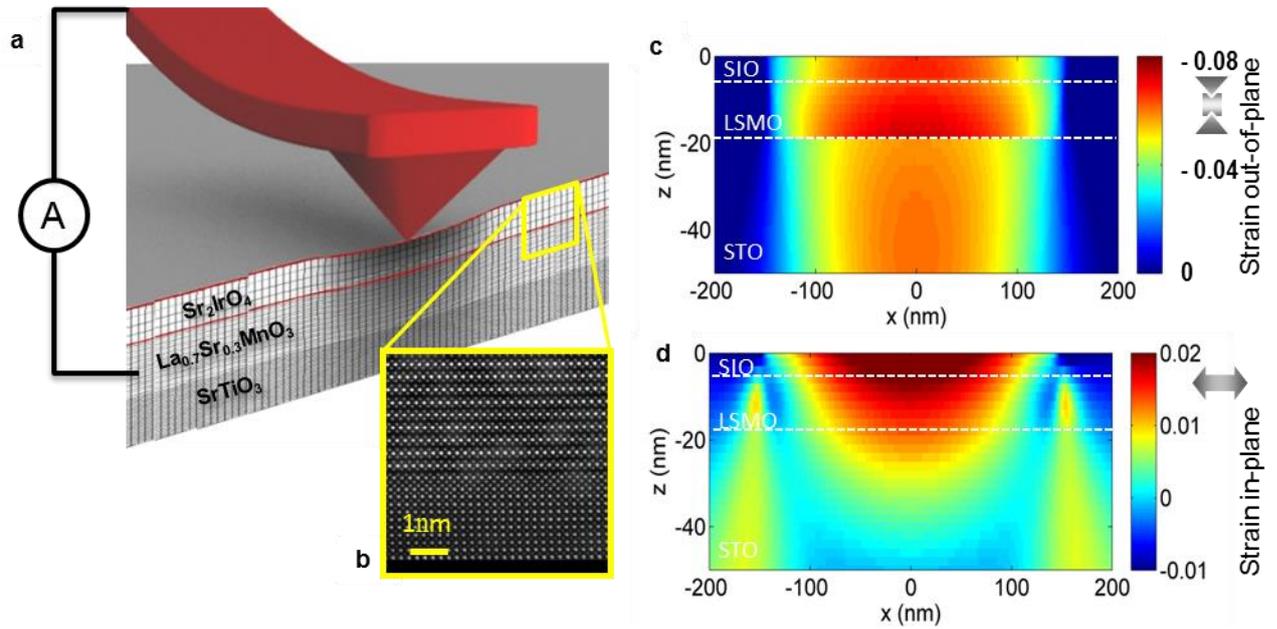

**Figure 1. a.** Schematic illustration of the experimental realization: a capacitor-like heterostructure is composed of the SIO layer placed between a conductive LSMO buffer layer acting as bottom electrode, and a conductive AFM tip that acts as both top electrode for current perpendicular to the plane (CPP) sensing and uniaxial pressure probe. **b.** High resolution scanning TEM of the SIO/LSMO interface **c.** Out-of-plane strain distribution, according to continuous elasticity models, of the heterostructure under tip-induced stress, considering a tip radius of 150 nm and 10 μN of applied force. **d.** Same type of calculation for the in-plane strain distribution.

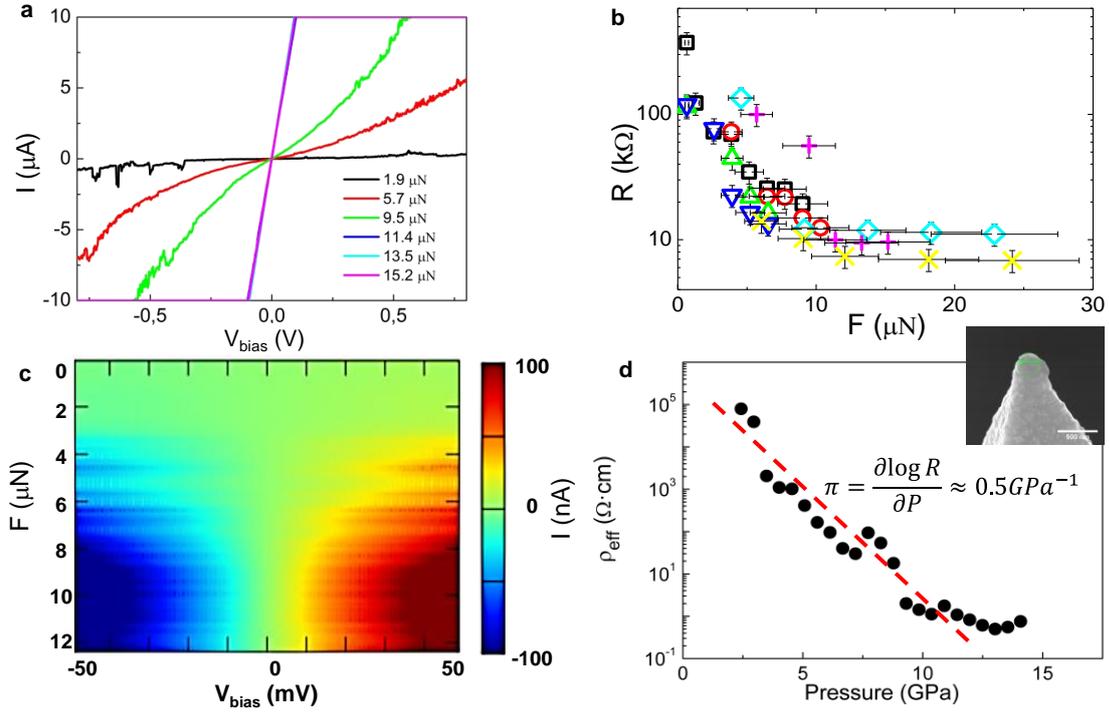

**Figure 2.** Local transport measurements **a.** I($V_{bias}$) characteristic curves obtained for applied loads ranging from 1.9 to 15.2 μN (Note that the three curves under higher loads overlap within this scale). **b.** Resistance as a function of the applied load. Data include different sets of experiments using PtIr coated tips ($r_{tip}$ ~50 nm) at a single point for diverse loads (black squares and red circles), at different points for each specific load (all the others), and at reduced relative humidity (pink crosses). **c.** 3D map of conductivity (color scale) as a function of tip-sample voltage ($V_{bias}$) and normal force (F), obtained with a B-doped diamond coated tip of $r_{tip}$ ~ 150 nm as determined by the SEM after the measurements (inset in d, bar scale is 500nm) **d.** Effective resistivity as a function of the applied pressure, both obtained from the 3D map. **c**. The red dashed lines are meant to guide the eyes. Note the evident logarithmic decay over pressure.

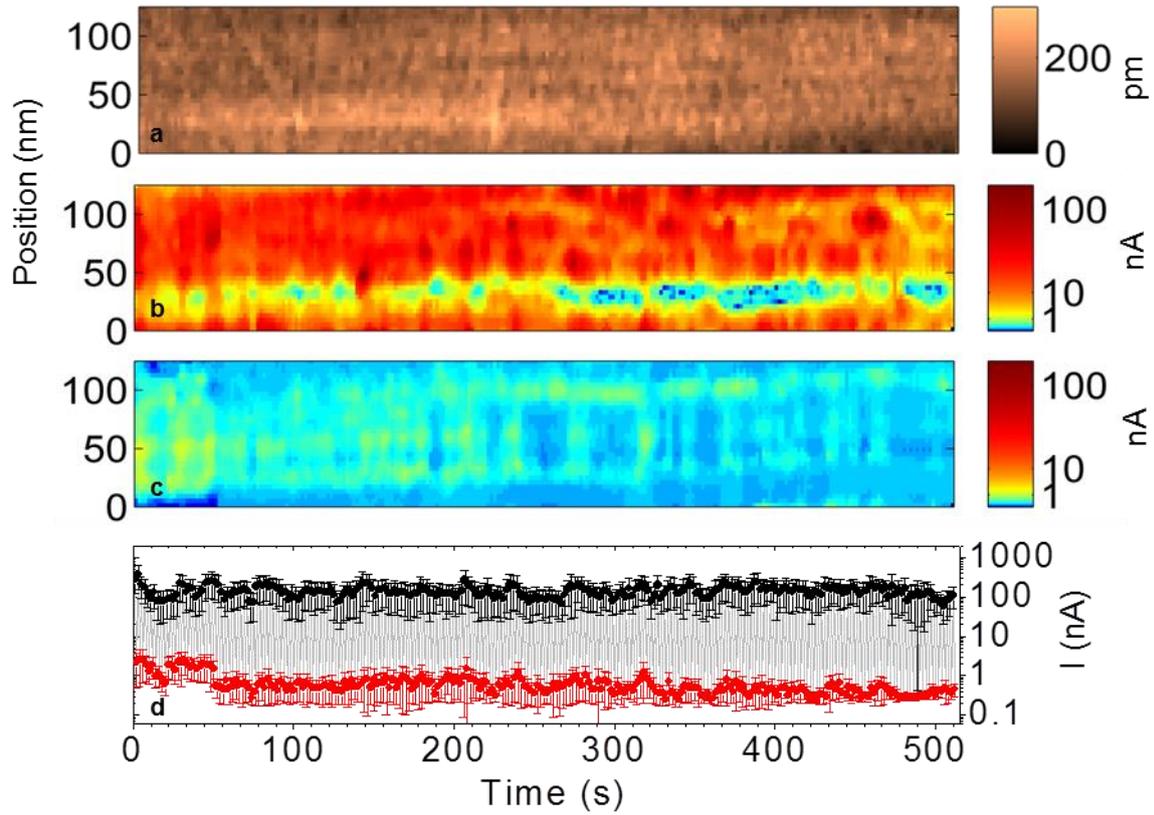

**Figure 3. a.** Topography monitoring of the 125 nm long line scanned over 512 times at a rate of 1Hz to continuously check the SIO film stability during the switch experiments. Successive scans are performed by increasing and releasing the force alternatively from 2 µN to 0.5 µN. The measured current is shown in **b** and **c,** respectively. The line averaged current for each load and scan are represented as points in **d** (black points for **b** and red points for **c**). The error bars correspond to the standard deviation. The experiment was performed using B-doped diamond coated tips.